\documentclass{article}
\usepackage{spconf,amsmath,graphicx}
\usepackage{comment}
\usepackage{marvosym}
\usepackage{xcolor}
\usepackage{hyperref}
\usepackage{soul}
\hypersetup{
    colorlinks=true,
    linkcolor=blue,
    filecolor=black,      
    urlcolor=blue,
    citecolor=blue
    }

\title{sVAD: A Robust, Low-Power, and Light-Weight \\ Voice Activity Detection with Spiking Neural Networks}
\name{Qu Yang$^{1*}$, Qianhui Liu$^{1}$\sthanks{Equal Contribution. \textsuperscript{\Letter} Corresponding author. qhliu@nus.edu.sg\\ This work is supported by IAF, A*STAR, SOITEC, NXP and National University of Singapore under FD-fAbrICS: Joint Lab for FD-SOI Always-on Intelligent \& Connected Systems (Award I2001E0053).}\textsuperscript{\Letter}, Nan Li$^{2}$, Meng Ge$^{1}$, Zeyang Song$^{1}$, Haizhou Li$^{3,1}$} 
\address{$^{1}$ National University of Singapore, Singapore \quad
      $^{2}$ Tianjin University, Tianjin, China \\
      $^{3}$ The Chinese University of Hong Kong, Shenzhen (CUHK-Shenzhen), China}
\begin{document}
%
\maketitle
\begin{abstract}
Speech applications are expected to be low-power and robust under noisy conditions. An effective Voice Activity Detection (VAD) front-end lowers the computational need. Spiking Neural Networks (SNNs) are known to be biologically plausible and power-efficient. 
However, SNN-based VADs have yet to achieve noise robustness and
often require large models for high performance. 
This paper introduces a novel SNN-based VAD model, referred to as sVAD, which features an auditory encoder with an SNN-based attention mechanism. Particularly, it provides effective auditory feature representation through SincNet and 1D convolution, and improves noise robustness with attention mechanisms. The classifier utilizes Spiking Recurrent Neural Networks (sRNN) to exploit temporal speech information. 
Experimental results demonstrate that our sVAD achieves remarkable noise robustness and meanwhile maintains low power consumption and a small footprint, making it a promising solution for real-world VAD applications.
\end{abstract}

\section{Introduction}
\label{sec:intro}
Voice Activity Detection (VAD) plays a pivotal role as a front-end in various speech applications such as automatic speech recognition, keyword spotting \cite{oh2019acoustic,price2017low,wu2021hurai}. It detects whether the speech is present in an audio signal to activate subsequent applications only when speech is detected. This paradigm helps to save computational resources and enhance the overall efficiency of the system \cite{yadav2022hardware}. 

VAD serves as an always-on model that is required to operate efficiently with low power consumption. Simultaneously, as a front-end, it needs to be light-weight, ensuring minimal memory utilization while maintaining its performance.
Beyond these prerequisites, noise robustness is also critical so that the VAD can operate stably under different noise levels. Hence, there is a demand for a robust, low-power, and light-weight VAD model.
 
Spiking Neural Networks (SNNs) mimic the information processing mechanism in the human brain \cite{liu2020effective, liu2022event,yang2022deep,yang2022training}. 
A spiking neuron is only active when it receives or emits a spike, facilitating power-efficient processing. Incoming spikes increment the neuron's membrane potential through accumulate (AC) operations, a process more power-efficient than the multiply-accumulate (MAC) operations in traditional artificial neural networks (ANNs) \cite{farabet2012comparison,Xinyin23NeurIPS,Gongfan23DepGraph,Xinyin23DeepCache,Xingyi22DeRy}. This intrinsic power-saving characteristic renders SNNs as optimal candidates for VAD applications.

However, SNN-based VAD suffers a significant performance loss under noisy conditions. It was reported that system performance in low speech-to-noise (SNR) ratio can be 25\% lower than that in high SNR conditions~\cite{dellaferrera2020bin}.
Additionally, achieving optimal performance often necessitates large SNN models. \cite{wu2021hurai} demonstrated the best SNN-based VAD results using a model containing over 1M parameters. Such a large model can be impractical in many real-world applications. \cite{dellaferrera2020bin} employed a compact 2.6K-parameter model, which regrettably shown poor performance. 

To address these limitations, we propose a novel SNN-based VAD model, referred to as sVAD, that is expected to be noise-robust, low-power, and light-weight. 
We present an auditory encoder integrated with an SNN-based attention mechanism. 
This encoder employs both SincNet \cite{ravanelli2018speaker} and 1D convolution to extract flexible and interpretable auditory features in a data-driven manner, thereby resulting in effective feature representation and enhancing the overall performance.
The incorporation of the SNN-based attention mechanism improves the saliency of the extracted features, thus increasing the robustness of VAD model. The classifier employs the Spiking Recurrent Neural Networks (sRNN) that can exploit the temporal information contained in the speech. Experimental results demonstrate our proposed sVAD achieves remarkable noise robustness and meanwhile maintains low power consumption and a small footprint.

The rest of the paper is organized as follows. Section 2 presents the SNN-based VAD incorporating with auditory attention. Section 3 reports our experiments and results. Finally, we conclude the work in Section 4.

\section{SNN-based VAD with Auditory Attention}
\label{sec:method}

\begin{figure}[t]
\centering
\includegraphics[scale=1.0]{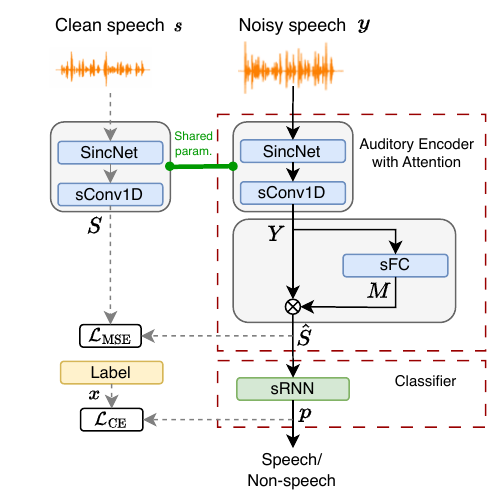}
\caption{The proposed SNN-based VAD model consists of an auditory encoder with attention for feature extraction and a classifier for frame-level classification.}
\label{fig:overview}
\end{figure}

As shown in Fig. \ref{fig:overview}, the proposed SNN-based VAD model 
\textcolor{black}{ consists of two key components: 1) an auditory encoder with attention, which converts raw audio input into spike-featured frames; and 2) a classifier dedicated to the frame-by-frame voice activity detection task.}
The VAD model benefits from the energy efficiency of event-driven computation of SNN.

\subsection{Spiking Neuron Model}
\label{sec:Spiking_Neuron}
We employ the widely recognized Leaky Integrate-and-Fire (LIF) model \cite{gerstner2002spiking} for spiking neurons. 
LIF neurons integrate synaptic inputs until their membrane potential exceeds a firing threshold, resulting in output spikes for transmission to subsequent neurons. The LIF neuron's dynamics are captured by the following discrete-time expressions:
\begin{equation}
	U_i^l[t] = \alpha U_i^l[t - 1] + I_i^l[t] - \vartheta O_i^l[t - 1],
	\label{eq: discrete_mem}
\end{equation}
\begin{equation}
	I_i^l[t] = \sum_j w_{ij}^{l-1} O_{j}^{l-1}[t-1] + b_i^l,
	\label{eq: input_current}
\end{equation}
where $U_i^l[t]$ represents the membrane potential of neuron $i$ at layer $l$, and $I_i^l[t]$ denotes the input current at time $t$. $\alpha$ signifies the membrane potential decay constant, while $\vartheta$ represents the neuronal firing threshold. $w_{ij}^{l-1}$ denotes the connection weight from neuron $j$ in the preceding layer $l-1$, and $b_i^l$ is the constant current injected into neuron $i$. The occurrence of an output spike, denoted as $O_i^l[t-1]$, is determined using the spiking activation function:
\begin{equation}
	O^l_i[t] = \Theta\left(U^l_i[t]-\vartheta\right).
	\label{eq: heaviside}
\end{equation}
Here, $\Theta(\cdot)$ is the Heaviside step function.

\subsection{Auditory Encoder with Attention}
\label{sec:encoder}

\subsubsection{Auditory encoder}
\label{sec:time-domian}
Our study leverages the well-established auditory encoder, SincNet \cite{ravanelli2018speaker}, known for its exceptional feature extraction capabilities. In contrast to conventional Convolutional Neural Networks (CNNs) processing raw speech data, SincNet introduces constraints on filter shapes in its first layer, making it highly suitable for our VAD model. It uses parametrized band-pass filters with only two parameters, ensuring interpretability and human intuition in the feature extraction process. By integrating SincNet into our auditory encoder, we enhance feature extraction, improving the performance of our SNN-based VAD model across various noise conditions while maintaining a light-weight configuration.

In general, 1D convolutional layers introduce trainable parameters enhancing encoder adaptability to data characteristics and improving performance through training.
Inspired by ConvTasNet \cite{luo2019conv}, we extend this data-driven ability by adding an SNN-based 1D convolutional layer (sConv1D) to SincNet-processed features. This empowers the encoder to optimize auditory feature extraction during training. 
Additionally, sConv1D outputs spikes, facilitating the transformation of raw audio into spike representation for subsequent SNN-based processing.

\subsubsection{Attention mechanism}
\label{sec:attention}
To enhance our VAD model robustness, especially in low SNR conditions, we introduce an attention mechanism inspired by the human auditory system's masking effect.
In the encoder module (shown in Fig. \ref{fig:overview}), we implement this attention mechanism as follows: Initially, we subject the extracted features $Y$ to three layers of SNN-based fully connected (sFC) layers to derive an attention mask $M$. Subsequently, we leverage this attention mask to modulate the extracted features, generating attended features denoted as $\hat{S}$:
\begin{equation}
	\hat{S} = Y \odot M
	\label{eq:attention}
\end{equation}
where $\odot$ is the element-wise multiplication. The trainable three-layer sFC block optimizes the attention mask for different acoustic conditions.

\subsection{Model Learning}
\label{sec:learning}
\subsubsection{Spiking recurrent neural network for classification}
\label{sec:srnn}
The classifier utilizes the encoded features as its input and is trained to perform frame-by-frame decision-making during runtime.
To enhance the sequential modeling capacity, we establish the classifier using SNN with recurrent connections, referred to as sRNN as shown in Fig. 1. Unlike the feedforward processing update formula (Eq. (\ref{eq: input_current})), the update equation for recurrent neurons includes an additional term for recurrent connections:
\begin{equation}
	I_i^l[t] = \sum_j w_{ij}^{l-1} O_{j}^{l-1}[t-1] + \sum_i w_{ii}^{l} O_{i}^{l}[t-1] + b_i^l, 
	\label{eq:recurrent}
\end{equation}
where $w_{ii}^{l}$ denotes the recurrent connection weight of neuron $i$ in layer $l$. 

\subsubsection{Loss function}
\label{sec:loss_func}
The proposed VAD model's loss function comprises two components: a classification loss and an attention mask loss.
For the classification loss, we employ cross-entropy (CE):
\begin{equation}
	\mathcal{L}_{\text{CE}} = -\sum^{2}_{c=1} x_c\log(p_c).
	\label{eq:ce_loss}
\end{equation}
Here, 
$[x_1, x_2]$ denotes an one-hot encoding ($[0,1]$ for speech and $[1,0]$ for non-speech), 
\textcolor{black}{and $p_c$ is the Softmax probability that the input belongs to class $c$.}
To calculate the attention mask loss, we utilize the mean squared error (MSE):
\begin{equation}
	\mathcal{L}_{\text{MSE}} = \frac{1}{N}\sum^N_{k=1} (\hat{S}_k - S_k)^2,
	\label{eq:mse_loss}
\end{equation}
where $S$ is the clean speech features, and $\hat{S}$ is the modulated features described in Eq. (\ref{eq:attention}). 
\textcolor{black}{$N$ denotes the total number of elements in the features and $k$ represents the index of elements.}
Consequently, the overall loss of our proposed sVAD model is defined as:
\begin{equation}
	\mathcal{L} = \mathcal{L}_{\text{CE}} + \lambda \mathcal{L}_{\text{MSE}}, 
	\label{eq:overall_loss}
\end{equation}
where $\lambda$ is the hyperparameter for balancing both losses.

\subsubsection{SNN training algorithm}
\label{sec:algorithm}
The stateful nature of spiking neurons, similar to vanilla RNNs, enables SNNs training by unfolding the network across all time steps and employing the Backpropagation Through Time (BPTT) algorithm \cite{werbos1990backpropagation}.
However, the non-differentiable spiking activation function, as outlined in Eq. \ref{eq: heaviside}, poses a challenge for direct BPTT application. 
This study addresses it by adopting the surrogate gradient approach \cite{neftci2019surrogate}, introducing a boxcar function as an effective surrogate gradient:
\begin{equation}
	\Theta' (U_i^{l}[t] -\vartheta) \approx \theta' (U_i^{l}[t] -\vartheta) =  \frac{1}{a}sign\left(|U_i^{l}[t]-\vartheta|<\frac{a}{2}\right),
	\label{eq: surrogate}
\end{equation}
where the hyperparameter $a$ controls the permissible range of membrane potentials that allow gradients to pass through.

\section{Experiments and Results}
\label{sec:experiments}

\subsection{Dataset and setup}
\label{sec:dataset_setup}
We evaluate the proposed sVAD model on the QUT-NOISE-TIMIT dataset \cite{dean2010qut}, consisting of $600$ hours of noisy speech. This dataset combines clean speech recordings from the TIMIT dataset with real-world noise scenarios, such as cafe, car, home, street, and reverberant environments. 
The dataset is categorized into three noise levels: low (SNR = +15dB, +10dB), medium (SNR = +5dB, 0dB), and high (SNR = -5dB, -10dB), with further stratification based on noise environment (Group A and Group B). Training and testing alternate between these groups, covering all noise levels. For the evaluation metrics, we report results using the \textcolor{black}{frame-level} Half-Total Error Rate (HTER), averaging the Miss Rate (MR) and False Alarm Rate (FAR). 

Feature extraction by the auditory encoder produces $20$-dimensional features with a $30ms$ frame size and $50\%$ overlap,
\textcolor{black}{enabling $15ms$ frame-by-frame classification and achieving a low latency of $15ms$ for our models.} The classifier comprises $32$ hidden recurrent spiking neurons and $2$ linear readout neurons for the ``speech'' and ``non-speech'' classes. Spiking neuron hyperparameters are set to $\alpha = 0.5$, $\vartheta = 0.3$, and $a = 4$.

All experiments use PyTorch with GPU acceleration, employing the Adam optimizer for 100 epochs, a batch size of 128, and an initial learning rate of 0.001. The learning rate reduces it by a factor of ten every 40 epochs.

\subsection{Results and analysis}
\label{sec:result_analysis}

\subsubsection{Compare with existing models}
\label{sec:comp}
We evaluate our sVAD model by comparing it with ten baseline systems, including standards-based approaches (e.g., ETSI \cite{doc2002speech} and G729B \cite{benyassine1997silence}), statistical methods (e.g., LTSD \cite{ramirez2004efficient} and Sohn \cite{sohn1999statistical}), machine learning-based solutions (e.g., GMM \cite{dean2010qut}, CLC \cite{ghaemmaghami2015complete}, and CNN \cite{silva2017exploring}), and recent SNN-based VAD models (e.g., Bin e. \cite{dellaferrera2020bin}, SNN h1/h1-p \cite{martinelli2020spiking}, and VAD system in HuRAI \cite{wu2021hurai}).

For a comprehensive and equitable comparison, we first assess all SNN-based VAD models, considering parameters count and HTER across different noise levels (low, medium, and high). Results are reported in Table \ref{tab:overall_result}. 
Notably, the VAD in HuRAI employs significantly more parameters, making it impractical for real-world use \cite{yadav2022hardware}. To ensure fairness, we re-conduct the VAD in HuRAI by aligning the parameter count with those used in our model.
For a comparison with the Bin e. model, which uses only $2.6$K parameters, we create a reduced version, named sVAD-S, with a smaller sRNN layer of just $10$ recurrent spiking neurons and two sFC layers for attention mask estimation. Despite these reductions, our modified model exhibits only a marginal increase in HTER across all three noise levels. In addition, it is worth mentioning that our models use the smallest frame shift (i.e., $15ms$) for the encoder among all SNN-based models (e.g., $20ms$ in \cite{dellaferrera2020bin}), resulting in the lowest latency among them.

\begin{figure}[t]
\centering
\includegraphics[width=1.0\linewidth]{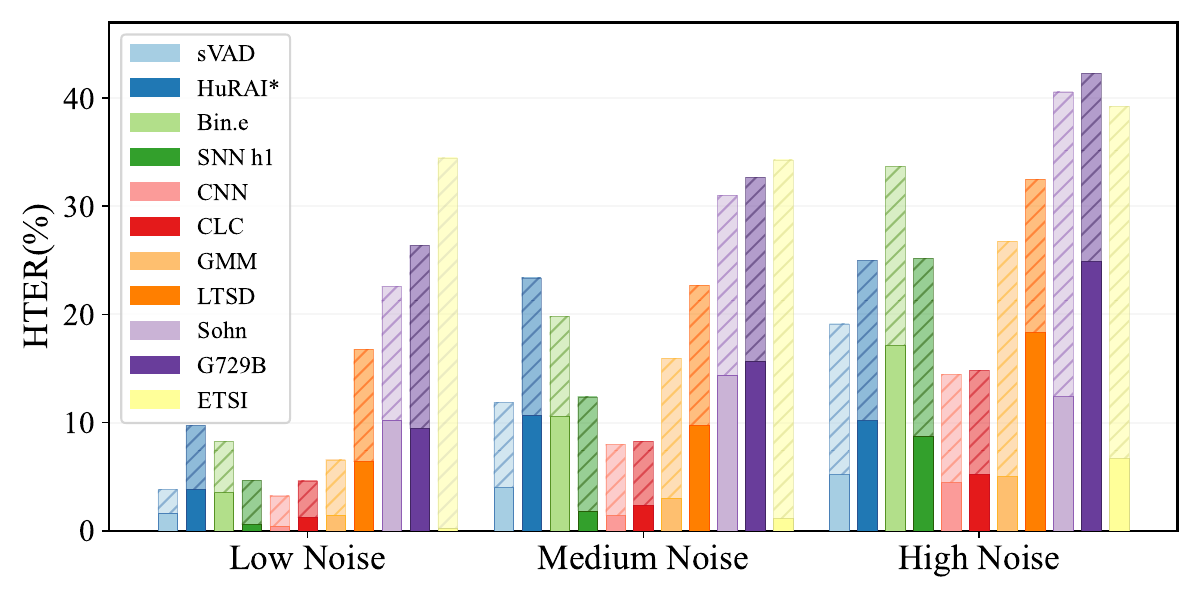}
\vspace{-0.3cm}
\caption{Comparisons of HTER performance for different noise levels. Dark and light shadings denote the miss rate and false alarm rate contributions, respectively.}
\label{fig:HTER_comp}
\vspace{-0.4cm}
\end{figure}

Next, we compare our sVAD model with the aforementioned ten baseline systems, as depicted in Fig. \ref{fig:HTER_comp}. 
The results demonstrate that our sVAD outperforms most models, with the exception of the CNN and CLC models. This can be attributed to the fact that the CNN and CLC models use larger frame sizes and shifts for decision-making (with frame sizes and shifts more than 5 times for CNN and more than 3 times for CLC in comparison to our models), thereby introducing increased latency.
Despite this discrepancy, it's crucial to emphasize that the primary focus of this study is the development of a robust, low-power, and light-weight SNN-based VAD model under noisy conditions rather than striving for the top HTER performance.

Overall, our model stands out among SNN-based VAD models, delivering superior performance, especially in high-noise conditions, while maintaining a competitive parameter count. This achievement signifies the successful development of a robust, low-power, and light-weight SNN-based VAD model under noisy conditions.

\begin{table}[t]
\centering	
\caption{\textcolor{black}{A summary of SNN-based VAD models, including parameter count and HTER across noise levels (Low, Medium, and High). * denotes our reproduced results.}}
\setlength\tabcolsep{3pt}
\resizebox{0.47\textwidth}{!}{%
\begin{tabular}{lccccc}
\hline
\textbf{Model}  & \textbf{\# Param.} & \textbf{Low (\%)} & \textbf{Medium (\%)} & \textbf{High (\%)} \\
\hline
HuRAI \cite{wu2021hurai}                   & $>$1M      & 2.7     & 6.7      & 15.0 \\
HuRAI* \cite{wu2021hurai}           & 4.4K      & 9.8     & 23.4     & 25.0 \\
Bin e. \cite{dellaferrera2020bin}         & 2.6K      & 8.2     & 23.6     & 33.6 \\
SNN h1 \cite{martinelli2020spiking}         & 26K      & 4.6   & 12.4     & 25.2 \\
SNN h1-p \cite{martinelli2020spiking}       & 4.1K      & 4.7     & 12.5    & 25.8\\
\textbf{sVAD}                    & 4.3K      & 4.0     & 11.9      & 19.1 \\
\textbf{sVAD-S}                  & 2.4K      & 5.8     & 12.6     & 22.3 \\
\hline
\end{tabular}}
\vspace{-0.25cm}
\label{tab:overall_result}
\end{table}

\begin{table}[!t]
\centering	
\caption{\textcolor{black}{Ablation study on the proposed Auditory Encoder's efficacy, focusing on high-noise conditions (SNR: -5dB and -10dB).}}
\begin{tabular}{lccc}
\hline
\textbf{Encoder Model}  & \textbf{MR} & \textbf{FAR} & \textbf{HTER} \\
\hline
sVAD                    & 10.4\%     & 27.9\%     & 19.1\% \\
- sConv1D               & 19.5\%     & 26.9\%     & 23.2\% \\
- sConv1D \& Attention  & 24.6\%     & 29.8\%     & 27.2\% \\
\hline
\end{tabular}
\vspace{-0.4cm}
\label{tab:ablation}
\end{table}
\subsubsection{Ablation study}
\label{sec:ablation}
To validate the efficacy of our proposed auditory encoder, we conduct an ablation study on high-noise conditions, with results detailed in Table \ref{tab:ablation}.  
We employ the proposed auditory encoder used in sVAD as the baseline.
First, we remove the learnable sConv1D layer from the encoder, resulting in a significant $4.1\%$ increase in HTER. Subsequently, the further elimination of the attention mechanism led to another noteworthy $4.0\%$ increase in HTER.
In summary, our ablation study demonstrates compelling evidence for the efficacy of our auditory encoder in significantly improving the robustness of our SNN-based VAD model under noisy conditions.

\subsubsection{Power consumption}
\label{sec:power}
\begin{table}[t]
\centering	
\caption{Power consumption of our sVAD and other baselines.}
\begin{tabular}{ll}
\hline
\textbf{Model}  & \textbf{Power Consumption} \\
\hline 
sVAD / sVAD-S   & 2.0 / 0.9$\mu$W (lower bound)   \\
Bin e. \cite{dellaferrera2020bin}     &   3.8 $\mu$W (lower bound)   \\
SNN h1-p \cite{martinelli2020spiking} & 25.1 $\mu$W  \\
Yang et al \cite{yang20181muw} & 1  $\mu$W \\
Price et al \cite{price2017low} & 22 $\mu$W \\
Meoni et al \cite{meoni2018low}  & 559 $\mu$W \\
\hline
\end{tabular}
\label{tab:power}
\vspace{-0.5cm}
\end{table}
This section estimates the power consumption of our sVAD models and compares them with that of other VADs. Our estimation is based on the Loihi neuromorphic chip \cite{davies2018loihi} and follows the estimation methodology in \cite{dellaferrera2020bin}, encompassing power consumption from synaptic operations and neuron updates.
As depicted in Table \ref{tab:power}, our sVADs demonstrate lower power consumption, particularly the smaller sVAD-S model. It is worth noting that we only estimate the lower bound based on Loihi, which is different from \cite{yang20181muw,price2017low,meoni2018low} that have run on ASIC chips.

\section{Conclusion}
\label{sec:Conclusion}
We develop a novel SNN-based VAD mode consisting of an auditory encoder with attention for feature extraction and an sRNN for classification. The auditory encoder with attention can provide an effective and robust feature representation of the raw audio, which enables us to deploy a light-weight SNN while still maintaining competitive performance. Furthermore, the model's small footprint, coupled with the energy-efficient characteristics of SNN, results in low power consumption. Comparison experimental results with other VADs show that our proposed sVAD has high noise robustness, low power consumption, and a small footprint. The ablation study further validates the effectiveness and robustness of our proposed auditory encoder with attention.




\newpage
\bibliographystyle{IEEEtran}
\small
\bibliography{refs}

\end{document}